# The Role of Electrical and Thermal Contact Resistance for Joule Breakdown of Single-Wall Carbon Nanotubes

Eric Pop

*Department of Electrical and Computer Engineering, Micro and Nanotechnology Laboratory*
*University of Illinois at Urbana-Champaign, Urbana IL 61801, U.S.A.*
E-mail: epop@uiuc.edu, Tel: +1 217 244-2070

Several data sets of electrical breakdown in air of single-wall carbon nanotubes (SWNTs) on insulating substrates are collected and analyzed. A *universal* scaling of the Joule breakdown power with nanotube length is found, which appears independent of the insulating substrates used or their thickness. This suggests the thermal resistances at the interface between SWNT and insulator, and between SWNT and electrodes, govern heat sinking from the nanotube. Analytical models for the breakdown power scaling are presented, providing an intuitive, physical understanding of the breakdown process. The electrical and thermal resistance at the electrode contacts limit the breakdown behavior for sub-micron SWNTs, the breakdown power scales linearly with length for microns-long tubes, and a minimum breakdown power (~ 0.05 mW) is observed for the intermediate (~ 0.5 μm) length range.

The electrical and thermal behavior of single-wall carbon nanotubes has been intensely studied in recent years [1-6]. Controlled electrical breakdown of SWNTs, i.e. "cutting," has been used to create electrodes for single-molecule experiments [7, 8], and their breakdown has been noted in the context of field-emission sources [9, 10] and bulk thermogravimetric (TGA) experiments [11, 12]. In-air electrical breakdown of SWNTs supported by an insulating substrate has also been proposed as a selective mechanism for preferentially eliminating metallic nanotubes among semiconducting ones as a bottom-up approach to building SWNT circuits [13, 14].

However, little is known specifically about SWNT device breakdown and reliability at high temperature, and about the role the ubiquitous contacts play in power generation and dissipation. Understanding and controlling the breakdown power (voltage) of nanotubes is also important for selective elimination of metallic among semiconducting SWNTs in electronic circuits. While previous work has developed electro-thermally coupled metallic nanotube transport models [1, 15], this manuscript analyzes, for the first time, the specific role played by the nanotube-electrode contacts in electrical and thermal transport and high-voltage breakdown.

SWNT breakdown voltage data is collected from studies by Seidel *et al.* [14] (from here on referred to as the Infineon data set), Maune *et al.* [16] (the Caltech data set), Javey *et al.* [17] and Pop *et al.* [15] (the Stanford data sets). The devices in these various studies share a similar geometric layout, i.e. a single-wall nanotube bridging two metallic contacts on top of an insulating material layer (Fig. 1a). The silicon wafer beneath is used as a back-gate, where necessary, to fully turn on the semiconducting tubes. The top of the nanotube is left uncovered and exposed to the ambient air. Only nanotubes whose complete *I-V* electrical characteristics were available, up to electrical breakdown, have been used in the present study. Combined, the set of electrically contacted SWNTs considered here covers a wide range of nanotube lengths (10 nm < $L$ < 8 µm), diameters (0.8 < $d$ < 3.2 nm), and electrical contact resistance (9 < $R_C$ < 830 kΩ). The physical dimensions are typically obtained from AFM measurements, while the electrical contact resistance ($R_C$ for the two contacts combined) is estimated from the linear portion of the *I-V* curve at low bias, and therefore incorporates the quantum contact resistance ($h/4q^2$ ≈ 6.5 kΩ).

The mechanism for in-air electrical breakdown of SWNTs is as follows. The voltage applied across the nanotube is raised (and the current typically increases) until the power dissipated is large enough to cause significant self-heating of the SWNT. If the power dissipation is uniform, the peak temperature occurs in the middle of the tube, and once this point reaches the breakdown temperature the nanotube oxidizes (burns) irreversibly. This yields a sharp drop to zero in the *I-V* curve, and a physical "cut" in the nanotube itself. A cartoon of this process is shown in Fig. 1a. Figure 1b displays the temperature profiles computed along a 3 µm nanotube at various voltages, using the model described in Ref. [15]. Breakdown voltages in vacuum or an inert ambient (e.g. Ar) are known to be significantly higher than those in air [7, 18], suggesting this is indeed an oxidation-induced breakdown [19]. In addition, AFM imaging of broken SWNTs [17] shows these cuts occur near the middle of the tube, where the temperature peaks. The breakdown temperature of SWNTs is known to be approximately $T_{BD}$ ≈ 600 °C from thermogravimetric (TGA) analysis of bulk samples [11, 12]. A range of ±100 °C around this value is generally accepted, somewhat dependent on the diameter (smaller diameter tubes are more reactive [20]) and impurities or defects present on the tubes.

Figures 2 and 3 show the experimental data gathered here, from the studies mentioned above. The data are shown both as breakdown voltage $V_{BD}$ vs. length (Fig. 2), and breakdown power $P_{BD}$ vs. length (Fig. 3). Figures 2b and 3b present a "zoom-in" of the data for the shorter nanotubes. Empty symbols represent the original raw data, whereas solid symbols represent the *intrinsic* breakdown power and voltage, after the power dissipated ($I_{BD}^2 R_C$) and voltage dropped ($I_{BD} R_C$) at the contacts were removed. Note the effect of removing $R_C$ from the breakdown data, which renders the trends of $V_{BD}$ and $P_{BD}$ scaling to appear more clearly.

The temperature profile $T(x)$ along the SWNT during Joule heating from current flow is given by the heat conduction equation:

$$A\nabla(k\nabla T) + p' - g(T - T_0) = 0 \qquad (1)$$

where $A = \pi db$ is the cross-sectional area ($b \approx 0.34$ nm the tube wall thickness), $k(T)$ is the SWNT thermal conductivity [21][22], $p' = I^2 dR/dx$ is the local Joule heating rate per unit length, $g$ is the heat loss rate to the substrate and ambient per unit length, and $T_0$ is the ambient temperature. An explicit solution of Eq. 1 above does not exist because both $k$ and $p'$ (through the resistance $R$) are functions of temperature, and therefore of position along the tube. A finite-difference numerical solution computed self-consistently with the electrical resistance was provided in Ref. [15]. However, practical approximations and meaningful insight into the breakdown conditions may be obtained by assuming an average thermal conductivity and power dissipation ($p' \approx P/L$) along the nanotube. This is the approach pursued here.

For *very long* tubes, an analytical solution of the peak temperature can be written, yielding a simple, linear expression describing the breakdown power [15]:

$$P_{BD} \approx g(T_{BD} - T_0)L, \qquad (2)$$

with the breakdown voltage being $V_{BD} = P_{BD}/I_{BD}$. The breakdown power $P_{BD}$ is the power input for which the peak temperature of the SWNT (in its middle) reaches the breakdown temperature $T_{BD}$. Eq. 2 is the "best-fit" straight dashed line in Figs. 2 and 3, with slope $g(T_{BD} - T_0) \approx 89$ W/m. Note this expression is independent of the thermal conductivity $k$, indicating that for long nanotubes the Joule heat is dissipated mostly down into the substrate, rather than laterally into their contacts (Fig. 1). This is in accord with the relatively flat temperature profiles calculated in Fig. 1b. With the assumption of $T_{BD} = 600 \pm 100\ ^\circ C$, Eq. 2 here yields a range for the heat sinking coefficient $g \approx 0.15 \pm 0.03$ WK$^{-1}$m$^{-1}$, consistent with the 0.17 value found in Ref. [15], but bearing in mind that the present study spans multiple data sets and a much wider range of SWNT diameters, substrates and lengths. This is significantly lower than the thermal conductance of radial (semi-cylindrical) heat flow into any of the insulating substrates here alone (SiO$_2$, Si$_3$N$_4$ or Al$_2$O$_3$), which is of the order 1 WK$^{-1}$m$^{-1}$ or greater, indicating that heat dissipation from the nanotube is limited by the nanotube-substrate interface [15].

At the other length extreme, the simple expression above cannot be used to describe the thermal and breakdown behavior of *very short* nanotubes (Figs. 2b and 3b). At first glance, an approximate solution of the heat conduction Eq. 2 in this length range would lead to

$$P_{BD} \approx (T_{BD} - T_0)\left(\frac{L}{8kA}\right)^{-1}, \qquad (3)$$

which predicts a $1/L$ dependence of the breakdown power. However, this implies an infinitely large breakdown power (and voltage) as the nanotube length approaches zero, which is evidently *not* observed experimentally. The key to understanding the experimental data is to realize there is a finite thermal resistance ($\mathcal{R}_T$) associated with each of the two nanotube-electrode contacts. This yields a finite temperature drop at each contact, locally given by $\Delta T_C = T_C - T_0 = kA\mathcal{R}_T |dT_C/dx|$, as shown in Fig. 1b. A more appropriate, yet simple expression of the breakdown power for very short nanotubes including this thermal contact resistance becomes

$$P_{BD} \approx (T_{BD} - T_0)\left(\frac{L}{8kA} + \frac{\mathcal{R}_T}{2}\right)^{-1}, \qquad (4)$$

which is the dash-dotted line in Fig. 3b, with $\mathcal{R}_T = 1.2 \times 10^7$ K/W. This value of the nanotube-electrode thermal contact resistance is consistent with typical metal-dielectric interface thermal resistance when normalized by the small contact area here [15, 23]. This gives a finite $P_{BD} \approx 0.1$ mW for the shortest tubes, as their length (electrode separation) approaches zero, as seen experimentally.

At this point it is relevant to inquire what the "long" and "short" length scales are for the applicability of the elementary approximations above. This can be better understood by writing down a less simple, yet still analytic solution of Eq. 1 which includes the thermal resistance at the contacts and covers the entire length range:

$$P_{BD} = gL(T_{BD} - T_0) \frac{\cosh(L/2L_H) + gL_H \mathcal{R}_T \sinh(L/2L_H)}{\cosh(L/2L_H) + gL_H \mathcal{R}_T \sinh(L/2L_H) - 1} \tag{5}$$

where $L_H = (kA/g)^{1/2} \approx 0.2$ µm is the characteristic thermal "healing" length along the SWNT. Note this reduces to Eqs. 2–3 in Ref. [15] when $\mathcal{R}_T = 0$, and down to Eqs. 2–4 above in the limits of very long and short nanotubes. This solution is plotted with the solid line in Fig. 3b, showing correct asymptotic behavior in the two length limits [24]. The "short" and "long" nanotube length range may now be thought of as compared to the order of the thermal healing length, $L_H$. Interestingly, these results suggest that the competing effects of heat sinking through the contacts vs. the substrate yield a *minimum breakdown power* (~0.05 mW) of electrically-heated SWNTs in air, for tubes with length in the range $2$–$3L_H \approx 0.4$–0.6 µm. Nanotubes much shorter than this break down at higher power inputs following Eq. 4 above, whereas longer nanotubes follow the simple linear trend of Eq. 2. For long SWNTs it appears acceptable to neglect the thermal resistance at the electrodes ($\mathcal{R}_T$) altogether, as other thermal conduction pathways become dominant: mostly, from nanotube down into substrate, with thermal resistance roughly equivalent to $1/gL$ (~ $2 \times 10^6$ K/W for a 3 µm SWNT, and less for longer nanotubes).

Before concluding, a number of issues must be commented on. In this simple analysis, the power dissipation ($p' = dP/dx \approx P/L$) and heat sinking ($g$) have been considered uniform along the nanotube. In an idealized scenario this is acceptable, as both are relatively weak functions of temperature and may be replaced with values averaged along the length of the SWNT. However, in practice they are both likely to depend of the nanotube-substrate separation. If even a small amount of "buckling" is present along the nanotube, the local thermal conductance into the substrate will be severely reduced, the local resistance (and power density) of the nanotube segment will be significantly higher, and local breakdown of the SWNT may be expected. Another possibility is that of a much enhanced electric field at a buckling or defect site. Theoretical simulations have recently shown that electrostatic breakdown can occur at localized fields of the order ~10 V/nm [25]. While these are much higher than the *average* axial fields caused by the lateral voltage in the data surveyed here (~10 V/µm), it is difficult to rule out such events at highly localized defect or buckle sites along the nanotubes. Both effects mentioned above are likely contributors to the variance seen experimentally among the different data sets for long nanotubes, even after the electrical contact effects are removed.

For short nanotubes, the probability of a buckling effect is proportionally lower, but the role of the contacts is more significant. In other words, the scatter among the breakdown data of very short SWNTs is explained by the inconsistent $\mathcal{R}_T$ of contacts among different nanotubes, and even between the two electrodes of a single nanotube. The latter scenario will shift the peak of the temperature profile away from the middle of the nanotube [26]. A change by a factor of two in $\mathcal{R}_T$ may alter the breakdown power of a very short SWNT by about 50% (see Eq. 4). In addition to the quality of the interface, $\mathcal{R}_T$ is also partly determined by the length of the nanotube-electrode overlap. This is usually difficult to control, and for values below a few $L_H$ (<

0.5 µm) the thermal contact resistance will be adversely affected. Beyond a few $L_H$ the length of this overlap is unimportant, given the exponential drop of the temperature profile within the electrodes [27]. Finally, the diameter of the various SWNTs considered in this study also likely plays a role in their breakdown characteristics. This role is difficult to quantify unless many samples with similar diameters are systematically analyzed, but undoubtedly it controls their reactivity (here with $O_2$) and their "footprint," i.e. their effective heat conductance into the substrate, $g$. Both of these diameter effects are likely convolved within the three experimental data sets shown in Figs. 2 and 3.

In conclusion, this work studies in-air electrical breakdown characteristics of substrate-supported single-wall carbon nanotubes. Several published data sets were analyzed, spanning a wide range of nanotube diameters, lengths and contact properties. Nevertheless, a few simple, universal scaling rules were found to emerge, showing that the breakdown power of long nanotubes scales linearly with their length (electrode separation), whereas the breakdown of very short nanotubes is almost entirely limited by their contact resistance. The data and model show a minimum in the electrical power required to break a single-wall carbon nanotube (0.05-0.1 mW), which may be tailored through careful contact (resistance) engineering. Simple scaling models for the breakdown voltage are presented, which aid in obtaining an intuitive, physical picture of the factors limiting electrical and thermal transport, and high-voltage breakdown in substrate-supported SWNTs.

**Acknowledgement.** I thank Juanita Kurtin and Ali Keshavarzi at Intel for interesting discussions, which stimulated my pursuit of this problem.

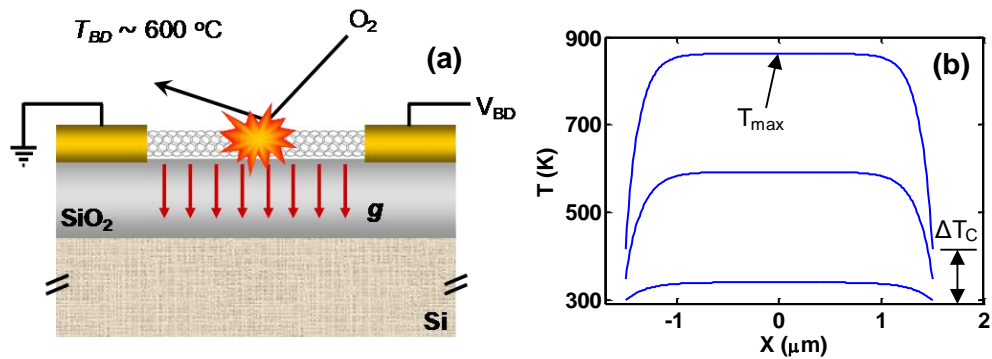

Fig. 1. (a) Schematic of oxidation-induced SWNT breakdown, when exposed to air at high applied voltage. The length of the nanotube portion between the contacts is $L$, the heat conductance into the substrate per unit length is $g$ (red arrows). (b) Calculated temperature profile for a 3 μm long tube at 3, 9 and 15 V bias from bottom to top. Note the peak temperature near the middle of the tube (where breakdowns are confirmed by AFM [17]) and the temperature drop $\Delta T_C$ at the contacts.

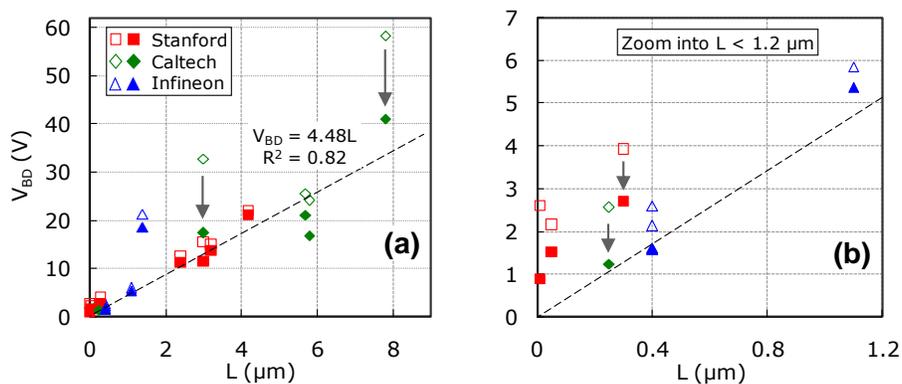

Fig. 2. (a) Breakdown voltage vs. SWNT length from the Stanford [16, 17], Caltech [15] and Infineon [14] data sets. Empty symbols are before, and solid symbols are after removing the electrical contact resistance drop $IR_C$ (arrows highlight some of the changes). (b) Same data sets, zoomed into the shorter nanotube range.

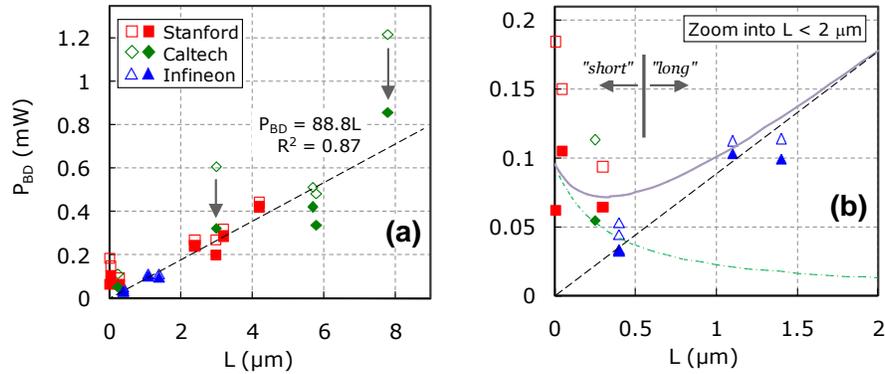

Fig. 3. (a) Breakdown power vs. SWNT length from the Stanford [16, 17], Caltech [15] and Infineon [14] data sets. Empty symbols are before, and solid symbols are after removing the contact power dissipation $I^2R_C$ (arrows highlight some of the changes). (b) Same data sets, zoomed into the shorter nanotube range. Dash-dot line is the short-nanotube approximation including $\mathcal{R}_T$ (Eq. 4), dashed line is the long-nanotube approximation (Eq. 2), and solid line is the solution spanning both ranges (Eq. 5). The finite breakdown power at near-zero length cannot be reproduced without including the SWNT-electrode contact thermal resistance $\mathcal{R}_T$.